\pdfoutput=1
\documentclass[usenatbib]{mn2e}
\usepackage{apjfonts}
\usepackage{amssymb}
\usepackage{amsmath}
\usepackage{ctable}
\usepackage{fixltx2e} 

\newcommand{\be}{\begin{equation}}
\newcommand{\ee}{\end{equation}}

\newcommand{\msun}{M_{\sun}}

\newcommand{\lstar}{L_{\ast}}

\newcommand{\scaleup}{}

\newcommand\plotone[1]
 {\centering \leavevmode \includegraphics[width={0.99\columnwidth}]{#1}}
\newcommand\plotonesmall[1]
 {\centering \leavevmode \includegraphics[width={0.85\columnwidth}]{#1}}

\newcommand{\acknowledgments}{\begin{small}\section*{Acknowledgments}\end{small}}
\newcommand\altaffilmark[1]{$^{#1}$}
\newcommand\altaffiltext[1]{$^{#1}$}
\title[Galaxy Cusps]{An Explanation for the 
Slopes of Stellar Cusps in Galaxy Spheroids\vspace{-0.5cm}}

\author[Hopkins and Quataert]{
\parbox[t]{\textwidth}{ 
Philip F. Hopkins\altaffilmark{1}\thanks{E-mail:phopkins@astro.berkeley.edu} 
\&\ Eliot Quataert\altaffilmark{1}} 
\vspace*{6pt} \\
\altaffiltext{1}{Department of Astronomy and Theoretical Astrophysics Centre, University of California
  Berkeley, Berkeley, CA 94720\vspace{-1.1cm}} }

\date{Submitted to MNRAS, September, 2010\vspace{-0.6cm}}
\begin{document}
\maketitle
\label{firstpage}

\begin{abstract}
The stellar surface mass density profiles at the centers of typical $\sim L_{\ast}$ 
and lower-mass spheroids exhibit power law 
``cusps'' with $\Sigma \propto R^{-\eta}$, where $0.5 \lesssim \eta \lesssim 1$
for radii $\sim 1-100$ pc.   Observations and theory support models in which these cusps are formed by dissipative gas inflows and nuclear starbursts in gas-rich mergers. At these comparatively large radii, stellar relaxation is unlikely to account for, or strongly modify, the cuspy stellar profiles.
We argue that the power-law surface density profiles observed are a natural consequence of the gravitational instabilities that dominate angular momentum transport in the gravitational potential
of a central massive black hole.
The dominant mode at these radii is an $m=1$ lopsided/eccentric disk instability, in which stars torquing the gas can drive rapid inflow and accretion.  Such a mode first generically appears at large radii and propagates inwards by exciting eccentricities 
at smaller and smaller radii, where $M_{\ast}(<R)\ll M_{\rm BH}$. 
When the stellar surface density profile is comparatively shallow with
$\eta<1/2$, the modes cannot efficiently propagate to 
$R=0$ and so gas piles up and star formation steepens the profile. 
But if the profile is steeper than $\eta = 1$, the inwards propagation of eccentricity 
is strongly damped, suppressing inflow and bringing $\eta$ down again. Together these results produce an equilibrium slope of $1/2\lesssim \eta \lesssim 1$ in the potential of the central black hole.   These physical arguments are supported by nonlinear numerical simulations of gas inflow in galactic nuclei.  Together, these results naturally explain the observed stellar density profiles of 
``cusp'' elliptical galaxies.
\end{abstract}

\begin{keywords}
galaxies: active --- galaxies: evolution --- 
quasars: general --- galaxies: nuclei --- galaxies: bulges --- cosmology: theory
\vspace{-1.0cm}
\end{keywords}

\vspace{-1.1cm}
\section{Introduction}
\label{sec:intro}

Observations have established that typical $\lesssim \lstar $ ellipticals and bulges exhibit 
steep central ``cusps'' in their surface luminosity density and stellar mass 
density profiles -- i.e.\ a continued rise in a
power-law like fashion towards small radii 
\citep{lauer91,lauer92:m32,crane93,
ferrarese:type12,kormendy94:review,lauer:95,kormendy99}. 
\citet{faber:ell.centers} showed that power-law nuclear profile 
ellipticals also tend to have higher degrees of rotational support 
and diskyness.  This, together with other observations \citep{kormendy99,quillen:00,
rest:01,lauer:bimodal.profiles,ferrarese:profiles,cote:smooth.transition}, 
has supported the idea that the cusp ellipticals are the direct 
product of gas-rich mergers and nuclear star formation during such mergers.  
Quantitatively,  power-law cusps have 
\be
I \propto \Sigma_{\ast} \propto R^{-\eta}
\ee
with $0.5\lesssim\eta\lesssim1$ representing the typical 
observed slopes; 
the power-law profile extends from the smallest radii observed 
in nearby spheroids ($\sim1\,$pc) to anywhere from 
$\sim10$ to $\sim 100$\,pc \citep{ferrarese:type12,jk:profiles}.\footnote{Provided 
it is defined over the same dynamic range, this is non-parametric and the same 
logarithmic slopes are recovered when Sersic fits are used.}
The most massive 
spheroids deviate from this behavior and exhibit flattened 
nuclear profiles, or ``cores.''  This is, however, widely 
believed to be due to ``scouring''  by a binary black hole in a gas-poor environment \citep[see e.g.][]{begelman:scouring} and thus does not reflect the initial formation history of the central stars that we focus on here.

\citet{barnes.hernquist.91} and \citet{mihos:starbursts.94} 
showed in simulations that tidal torques in mergers can drive rapid gas inflows, 
providing the fuel to power intense nuclear starbursts and build up the 
central stellar surface densities
\citep[see also][]{kormendy99,hopkins:cusps.ell}. 
This is observed in local Ultraluminous Infrared Galaxies (ULIRGs), whose nuclei constitute the most rapidly star-forming environments in the local Universe.   Moreover,
the observed central gas densities and star formation rates in ULIRGs will leave them with typical power-law like cusps when the starburst is complete \citep{hibbard.yun:excess.light,
tacconi:ulirgs.sb.profiles}.

Despite this progress, no theoretical explanation exists for {\em why} spheroid cusps 
should have a power law-like form in the range observed. 
The scales of the observed cusps are comparable to, or less than, the black hole (BH) radius of influence, and the potential is thus quasi-Keplerian.  At these radii, stars almost certainly formed primarily 
dissipatively in a gas-rich disk, rather than via violent relaxation of a pre-existing stellar disk. 
On the very smallest scales, two-body relaxation is expected to drive the system to  
a \citet{bahcallwolf} cusp; however outside of $\sim1\,$pc the relaxation time 
is $\gg t_{\rm Hubble}$. Moreover, the fact that the observed central cusps 
are disky and elliptical, often with significant rotational support (see references above) 
suggests that two-body relaxation has not had a large effect.   Instead, an understanding of the observed stellar cusps appears to require the combined effects of angular momentum redistribution and star formation in disky, gas+stellar+BH systems.  

Recently, we have shown that the formation of lopsided, eccentric disks within the BH radius of influence is a ubiquitous feature in hydrodynamic simulations of massive gas inflows 
in galaxies; such lopsided disks lead to efficient angular momentum 
transfer from the gas to the stars, powering BH accretion rates of up to $\sim10\,\msun\,{\rm yr^{-1}}$ \citep{hopkins:zoom.sims}. 
Moreover, the stellar relics of these disks are reasonably similar to the
nuclear disks observed on $\lesssim10\,$pc scales around 
nearby supermassive BHs \citep{hopkins:m31.disk}, particularly the well-studied case at the center of 
M31 \citep{lauer93}.   There are also many candidate nuclear disks observed in other systems \citep{lauer:ngc4486b,
  lauer:centers,
  houghton:ngc1399.nuclear.disk,
  thatte:m83.double.nucleus,debattista:vcc128.binary.nucleus,
  afanasiev:2002.ngc5055.nuclear.disk,
  seth:ngc404.nuclear.disk,ledo:2010.nuclear.disk.compilation}.

\begin{figure}
    \centering
    \scaleup
    \plotone{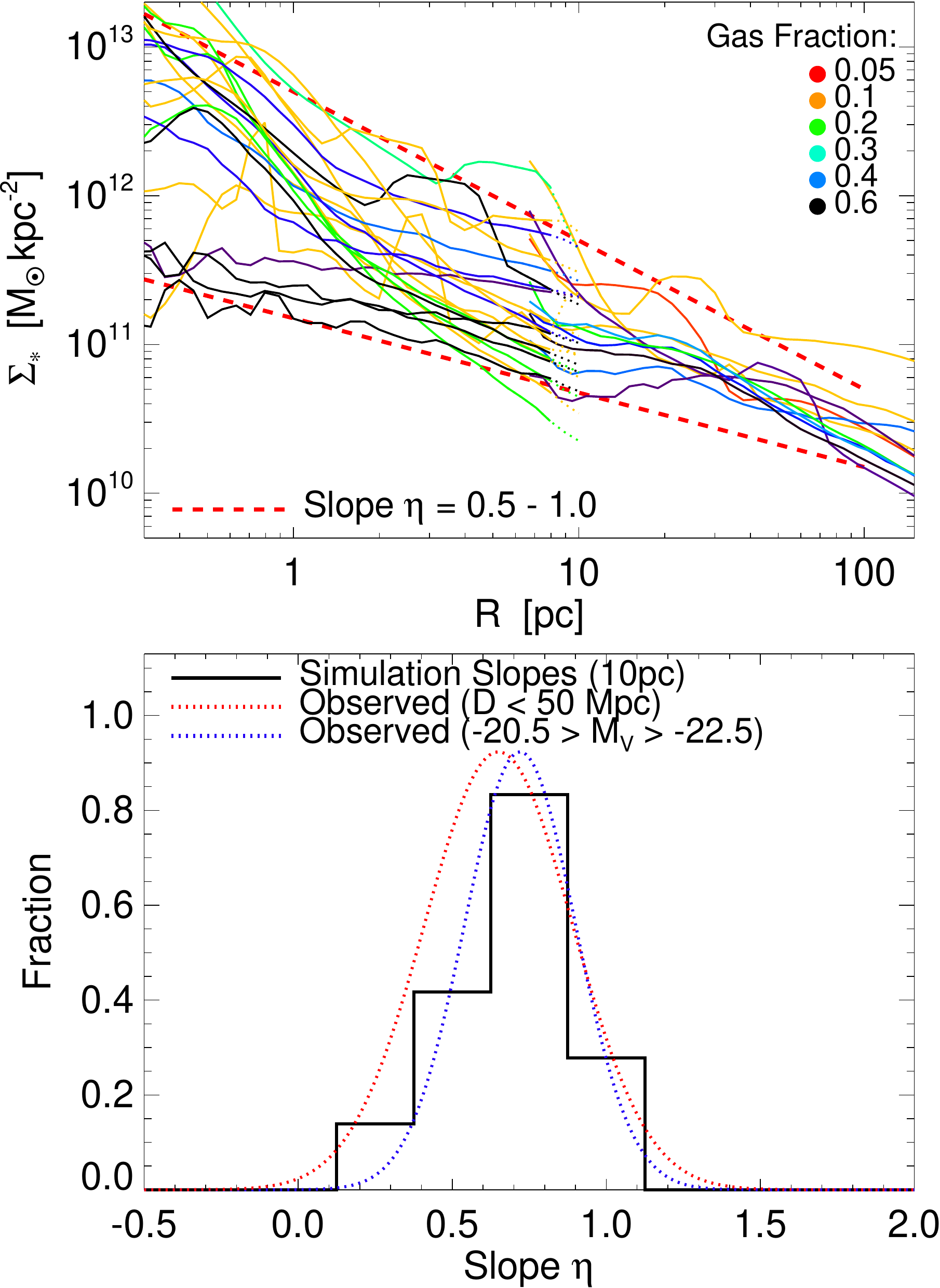}
    \caption{{\em Top:} Nuclear stellar mass profiles (cusps) produced in hydrodynamic 
    simulations of gas inflow driven by gravitational instabilities in dissipative starbursts;
    a lopsided/eccentric disk mode dominates at $\lesssim 100\,$pc.   Simulations of 
    different resolution/spatial scale are each shown only over the modeled dynamic 
    range; the highest resolution simulations extend to $\sim 100$ pc while more typical simulations begin at $\sim 10$ pc.   The simulations, taken 
    from \citet{hopkins:zoom.sims}, have a wide range of initial mass profiles, 
    gas fractions, and stellar feedback models. 
    However, in all cases the simulated mass profiles are consistent with cusps having
    power law-like slopes in the range $0.5\lesssim \eta \lesssim1$ 
    ($\Sigma\propto R^{-\eta}$; dashed line shows these 
    two cases, which roughly bracket the simulations). 
    {\em Bottom:} Distribution of cusp slopes.  The slopes are measured in each simulation 
    at $10\,$pc (fitting a power law to each from $3-30\,$pc or the maximum dynamic 
    range allowed within this interval), and from the observations in \citet{lauer:bimodal.profiles} 
    at approximately the same radius in both volume-limited and 
    magnitude-limited sub-samples (showing the ``cusp,'' as opposed to ``core'' 
    component of the observed bimodal distribution). 
    \label{fig:cusp.examples}}
\end{figure}

Figure~\ref{fig:cusp.examples} (top panel) shows the stellar surface density profiles at the end of the "nuclear-scale" and ``ultra-high'' resolution 
simulations of \citet{hopkins:zoom.sims}, which extend inwards from $\gtrsim10-100\,$pc with 
$\sim0.1\,$pc resolution; we show results in the quasi steady state phase of all simulations with significant inflows, $\ge 0.3\,\msun\,{\rm yr^{-1}}$ into $<1\,$pc, sustained for $>10^{5}\,$yr.  These SPH simulations include gas, stars, star formation, and a black hole as an additional collisionless particle; the simulations are idealized problems focused on studying the nonlinear evolution of gravitationally unstable systems in the potential of a massive black hole.  \citet{hopkins:inflow.analytics} show that the central dynamics and 
inflows are dominated by the nuclear $m=1$ modes. 
In Figure~\ref{fig:cusp.examples} the absolute 
stellar mass densities depend on the initial conditions (e.g., total gas mass), but the slopes are more robust; the simulations 
shown span a wide range in initial gas fractions, prescriptions for star 
formation and gas physics, 
initial stellar and gas mass profiles, and bulge-to-disk ratios 
(see Tables~1-3 in \citealt{hopkins:zoom.sims}), but converge to similar slopes. 
Comparing with the observed power law slopes of ellipticals (bottom panel), the agreement is 
reasonable. In this {\em Letter}, we provide a physical explanation for these results.

\vspace{-0.7cm}
\section{Propagation of Instabilities}
\label{sec:propagation}

Physically, the lopsided or eccentric disk mode (azimuthal wavenumber $m=1$ or amplitude $\propto \cos{\phi}$) is unique in any nearly Keplerian potential \citep{tremaine:slow.keplerian.modes}. Gravitational torques from other modes are suppressed by the gravity of the BH.   However, the resonant response between the epicyclic and orbital frequencies allows for global, low frequency
$m=1$ modes that can exert strong torques on the gas by inducing orbit crossing and shocks (e.g., \citealt{chang:m31.eccentric.disk.model, hopkins:inflow.analytics}).  Because of the importance of the $m = 1$ modes for redistributing gas inside the potential of the BH, we now focus on the physics of these $m = 1$ modes, in particular their propagation to smaller radii.

\vspace{-0.4cm}
\subsection{The WKB Limit}
\label{sec:wkb}

Consider an initially axisymmetric, thin, planar disk (surface density $\Sigma$) with a BH of mass $M_{\rm BH}$ at the coordinate center; we use cylindrical coordinates throughout ($R$, $\phi$, $z$).  The initial potential in the disk plane can be 
written $\Phi_{0} = \Phi_{0}(R)$, and other properties 
are defined in standard terms: 
\begin{align}
V_{c}^{2} & = R\, \frac{\partial \Phi }{\partial R} 
\approx \frac{G\,M_{\rm enc}(<R)}{R} \\ 
\Omega & \equiv  t_{\rm dyn}^{-1} = V_{c}/R \\ 
\kappa^{2} &\equiv  R\,\frac{{\rm d} \Omega^{2}}{{\rm d} R} + 4\,\Omega^{2} = 
\frac{\partial^{2}\Phi}{\partial R^{2}}+3\,\Omega^{2}
\end{align}
where $V_{c}$ is the circular velocity, $\Omega$ the angular 
velocity, and $\kappa$ the epicyclic frequency. 
We use $c_{s}$ to denote the sound speed in a gaseous 
disk and $\sigma_{z}$ the vertical dispersion in a stellar disk.

We consider a linear perturbation 
$\Sigma \rightarrow \Sigma_{0}(R) + \Sigma_{1}(R,\,\phi)$ 
(where $\Sigma$ is the total gas+stellar disk surface density) 
in a frame rotating with the perturbation pattern 
speed $\Omega_{p}$, and decompose the perturbation into 
linearly independent modes: 
\begin{align}
\Sigma_{m} &\equiv \Sigma_{a}(R)\,\exp{\left\{ i\,(m\,\phi-\omega\,t)\right\}} \\ 
\Sigma_{a}(R) &\equiv |a(R)|\,\Sigma_{0}(R)\,\exp{\left\{i\,\int^{R}\,k\,dR \right\}}
\end{align}
where $m$ is the azimuthal wavenumber, $|a|=|a(R)|$ the effective mode amplitude, 
$k$ the radial wavenumber, and the complex $\omega$ the mode 
frequency. With these definitions, the mode pattern speed is 
$\Omega_{p}\equiv {\rm Re}(\omega)/m$, and the mode 
growth rate $\gamma\equiv {\rm Im}(\omega)$.

We adopt a power-law disk as a convenient reference model: 
\be
\Sigma\propto R^{-\eta} = \Sigma_{0}\,\left(\frac{R}{R_{0}}\right)^{-\eta} \label{powerlaw}
\ee
It is straightforward to show then that 
\begin{align}
\Omega^{2} &=\frac{G\,M_{\rm BH}}{r^{3}}+
\frac{2\pi\,\alpha\,G\,\Sigma_{0}}{R_{0}}\,\left( \frac{R}{R_{0}} \right)^{-(\eta+1)} 
\end{align}
where 
$\alpha =(
\Gamma[1-\frac{\eta}{2}]\,\Gamma[\frac{1+\eta}{2}]
)/(
\Gamma[\frac{3-\eta}{2}]\,\Gamma[\frac{\eta}{2}]
)$ for $0<\eta<2$. 

We first consider modes in the WKB limit of tight-winding (i.e.\ local modes), where $|kR|\gg m$. We caution that this limit does not, in fact, 
hold for many of the global modes of most interest, but it is nevertheless 
instructive. We follow \citet{tremaine:slow.keplerian.modes}'s derivation for
a slow mode ($\Omega_{p}\ll \Omega$) 
in which the non-Keplerian part of the potential is  
small, i.e.\  
$\Phi = \Phi_{\rm BH} + \Phi_{d}$ where 
$\Phi_{d}/\Phi_{\rm BH} \sim M_{\rm d}/M_{\rm BH}\ll 1$.  Expanding
the equations of motion in terms $\mathcal{O}(\Phi_{d}/\Phi_{\rm BH})$ gives the WKB dispersion relation (to leading order in $|kR|^{-1}$) of 
quasi-Keplerian slow modes, 
\be 
\omega = \varomega + \pi\,G\,\Sigma_{d}\,|k|\,\Omega^{-1} 
- c_{s}^{2}\,k^{2}\,\Omega^{-1}
\label{eqn:slowmode.dispersion.gas}
\ee
for a gas disk, 
or 
\begin{align}
\omega &= \varomega + \pi\,G\,\Sigma_{d}\,|k|\,\Omega^{-1} \mathcal{F} \nonumber \\
&\approx \varomega + \pi\,G\,\Sigma_{d}\,|k|\,\Omega^{-1} \,\exp{\left(-\beta\,|kR| \right)}
\label{eqn:slowmode.dispersion.stellar}
\end{align}
for a stellar disk, 
where we define 
\begin{align}
\nonumber   
\varomega &\equiv \frac{\Omega^{2}-\kappa^{2}}{2\,\Omega} 
= - \frac{1}{2\,\Omega}\,\left( \frac{2}{r}\,\frac{d}{dr} + \frac{d^{2}}{dr^{2}} \right)\Phi_{d}\ .
\end{align}
In the dispersion relation for a stellar disk (eq. \ref{eqn:slowmode.dispersion.stellar}), $\mathcal{F}$ is the standard reduction factor \citep{binneytremaine}, and the latter equality in equation \ref{eqn:slowmode.dispersion.stellar} is a convenient approximation for softened gravity, with $\beta \approx \sigma_{z}/V_{c}\approx h/R$ (the stellar disk scale height). 

The $m = 1$ slow modes are stable in the limit $M_d \ll M_{\rm BH}$ \citep{tremaine:slow.keplerian.modes}.   Because of this physical constraint, the $m=1$ modes first appear at large radii -- the radius $\equiv R_{\rm crit}$ where $M_{d}/M_{\rm BH}\sim1$,  i.e.\ where the potential is transitioning to Keplerian \citep{hopkins:zoom.sims,hopkins:slow.modes}. 
The pattern speed $\Omega_{p}$ of the unstable mode is $\sim \Omega(R_{\rm crit})$. 
But if the mode can propagate inwards at constant $\Omega_{p}$, 
it will eventually be a slow mode, relative to the local $\Omega$ at smaller radii.

How does this propagation occur?  The wave packets propagate with 
approximate group velocity  $\sim V_{c}(R_{\rm crit})$,
so the timescale for the mode to travel is just the 
dynamical time at $R_{\rm crit}$. 
However, if these modes are forming in realistic ``initial'' disks, and if they are the dominant source of angular momentum transport, then the initial disk 
surface density profile cannot already be steep.   Dimensionally, $\varomega \sim (1/R\,\Omega)d\Phi_{e}/dR \sim 
G\,\Sigma/\Omega\,R$ (exactly true if the 
disk has a locally power-law profile; 
$\varomega = 
-\alpha\,(2-\eta)\,\pi\,G\,\Sigma/\Omega\,R$). 
Since $\Omega$ diverges $\propto r^{-3/2}$ at small radii, 
if the surface density profile is sufficiently shallow (and the 
dispersion is finite) then the 
right-hand side of Equation~\ref{eqn:slowmode.dispersion.gas} becomes 
arbitrarily small as $r\rightarrow0$, and finite $\omega$ cannot be supported -- 
the wave will refract back at some minimum radius $R_{\rm min}$. 
From Equation~\ref{eqn:slowmode.dispersion.gas}, this $Q$-barrier occurs when 
\be
|\Omega_{p}| \ge \varomega + \frac{\pi^{2}}{4}\,\frac{(G\,\Sigma_{d})^{2}}{c_{s}^{2}\,\Omega}
\label{eqn:omega.rcrit.limit}
\ee
in gas or 
\be
|\Omega_{p}|\ge \varomega + 
\frac{\pi\,G\,\Sigma}{e\,\beta\,R\,\Omega} \approx 
 \varomega + \frac{\pi\,G\,\Sigma}{e\,\sigma_{z}} 
\label{eqn:omega.rcrit.limit.stellar}
\ee
in stars. 
Since $\varomega \sim G\,\Sigma/\Omega\,R \propto R^{1/2-\eta}$ 
at small radii, for systems with finite $c_{s}$ or constant $\beta$ and a shallow 
$\Sigma \propto R^{-\eta}$ with $\eta \lesssim 1/2$, 
the initial waves cannot reach $R=0$.  Note that \citet{ostriker:eccentric.waves.via.forcing} 
show that the same restriction applies for modes in a pure fluid disk with a hard outer edge.

If the slope is too shallow, but the $m = 1$ modes are present, they will drive gaseous inflows 
that will ``pile up'' near the refraction radius.   
This will steepen the mass profile and increase the self-gravity at this radius, eventually allowing 
further mode propagation (in both gas and stars). Once $\eta\gtrsim1/2$,
then the RHS of Equation~\ref{eqn:slowmode.dispersion.gas} 
no longer vanishes as $r\rightarrow0$, and the 
modes can propagate through to $R=0$. Physically, the propagation can be understood as the eccentric mode at larger radii exciting strong eccentric perturbations at smaller radii. 

Consider two nearly-adjacent annuli at radii $R'$ and $R_{1}$: 
the material at $R_{1}$ is part of the $m=1$ mode, the material at $\le R'$ 
remains unperturbed. In the WKB limit the mode behavior at larger radii 
is swamped by the nearest asymmetric term -- i.e.\ just inside $R_{1}$, 
the perturbing potential is $\approx \Phi_{1}(R_{1})=2\pi\,G\,\Sigma_{1}(R_{1})\,|k|^{-1}$ 
(since there is no local 
corrugation to cancel this out). In the global limit the result is similar:  
at small radii inside an eccentric ring at radius $R_1$ having mass $M_{\rm ring}$ and $m = 1$ amplitude $|a|$, the magnitude of the local 
perturbed potential is just $\approx |a|\,G\,M_{\rm ring}/R_{1}\sim \pi\,G\,\Sigma_{1}(R_{1})\,R_{1}$, i.e., the same as the WKB result  with $|k|\sim R_{1}^{-1}$.
To estimate the velocity induced at smaller radii by this perturbed potential, we note that for a cold gas or stellar disk, the local pattern speed of the $m = 1$ mode is just 
$\omega-\varomega\approx\pi\,G\,\Sigma_{0}\,|k|\,\Omega^{-1}$.
Together, this leads to the result that the response (in both gas and stars) at smaller radii $\sim R'$ to the eccentric disk at larger radii is given by:
$|v_r/V_c| \sim (\Sigma_1/\Sigma_0) |kR|^{-1} \sim |a|$.
In other words, for a non-negligible mode amplitude 
$|a|\sim\Sigma_{1}/\Sigma_{0}$ and a global mode with $|kR|\sim 1$, large eccentricities and 
hence large coherent $m=1$ mode amplitudes, can be induced. 
The induced modes at these somewhat smaller radii can, in turn, excite large coherent eccentricities in 
the material at yet smaller radii, and so on, allowing the perturbation to grow
even at arbitrarily small $R$. 

The above derivation also implies, however, that there is a regime 
in which the inwards propagation of eccentricity will be inefficient. 
For a global mode at $R_{1}$, the 
perturbed potential is $\sim \pi\,G\,\Sigma_{1}(R_{1})\,R_{1}$, 
so the response $|e| \propto \Sigma_{1}(R_{1})\,R_{1} / \Sigma(R)\,R = 
|a(R_{1})|\,\Sigma(R_{1})\,R_{1}/\Sigma(R)\,R$. 
For a sufficiently flat mass profile 
$\Sigma(R_{1})\,R_{1} > \Sigma(R)\,R$, i.e., $\eta < 1$ for
$\Sigma(R)\propto R^{-\eta}$, the induced perturbation is 
large down to arbitrarily small $R$. But if the mass profile is 
too steep, $\Sigma(R_{1})\,R_{1} < \Sigma(R)\,R$, or $\eta>1$, 
then although the mode can formally be supported even at small 
radii, the induced amplitude will decline as one 
moves to $R\rightarrow0$. 
Crudely, we expect the propagation efficiency defined as 
$\log(|a(R)|/|a(R_{1})|)$ to decline $\propto-\eta$, for $\eta>1$. 
This is explicitly demonstrated for a very large sample of 
models by \citet{zakamska:eccentricity.wave.propagation} in the context of eccentricity propagation from an external perturber to the inner planets in planetary systems (the cutoff at $\eta<1/2$ is not evident in this approach, however, because of the discrete nature of the problem).    If $\eta > 1$ is established by some means, the low efficiency of eccentricity propagation to small radii implies that gas will not inflow as efficiently at small radii.   This will flatten the gas density profile and star formation will do the same for the stellar density profile, providing a mechanism for the system to self-adjust to have  $\eta \lesssim 1$.

\vspace{-0.5cm}
\subsection{Linear Global Modes}
\label{sec:exact}

The WKB results above are not exact, especially when the modes of interest are global and 
the disk mass is significant relative to $M_{\rm BH}$ (both of which are typically the case!).    To show that our conclusions are robust, we also demonstrate the same points regarding the propagation of modes using exact linear solutions 
for particular global normal modes.   Our methodology is described in detail in \citet{hopkins:slow.modes}, which we briefly summarize here.   We define $R_{0}$ for the power-law disk model  (eq. \ref{powerlaw})
so that $M_{d}(<R_{0})=M_{\rm BH}/[\alpha\,(2-\eta)]$ -- 
i.e.\ $\Sigma_{0}=M_{\rm BH}/(2\pi\,\alpha\,R_{0}^{2})$.  Unlike in the WBK analysis we do not expand the equations to linear order in $M_d/M_{\rm BH}$ but keep the full linear perturbation equations.
We consider a stellar-dominated (collisionless) disk since in our numerical results, the dominant torques in the gas are due to the stellar modes.    The resulting equations of motion for linear perturbations  are
\begin{align}
\nonumber 0 = & -(\Omega-\omega)\,\Sigma_{a}  \\ 
\nonumber & 
+\frac{\Sigma}{r^{2}\,\Delta}\,\left[ 2\,\Omega\,(\nu_{\Sigma}+\nu_{\Omega}-\nu_{\Delta}) - (\Omega-\omega)
\right]\,\Phi_{a}  \\ 
\nonumber & 
+\frac{\Sigma}{r\,\Delta}\,\left[ (\Omega-\omega)\,(1+\nu_{\Sigma}-\nu_{\Delta}) + \Omega\,(2+\nu_{\Omega}) - 
\frac{\kappa^{2}}{2\,\Omega} \right]\,\Phi_{a}^{\prime}  \\ 
& 
+\frac{\Sigma}{\Delta}\,\left[ \Omega-\omega\right]\,\Phi_{a}^{\prime\prime}
\label{eqn:mode.eom}
\end{align}
where $\Phi_{a} = \int_{0}^{\infty}\,{dr^{\prime}}\,{r^{\prime}}\,P(r,\,r^{\prime})\,\Sigma_{a}(r^{\prime})$ 
follows from Poisson's equation,\footnote{
The kernel $P$ is defined by
$P(r,\,r^{\prime}) = 
-{\pi\,G}\,b_{1/2}(r_{<}/r_{>})/{r_{>}} + {\pi\,G\,r}/{r^{\prime\ 2}}$
and includes the direct 
and indirect components of the potential, respectively. 
The Laplace coefficient $b_{1/2}$ is given by
\be
b_{1/2}(x)=\frac{2}{\pi}\int_{0}^{\pi}\,\frac{\cos{\theta}\,{d\theta}}
{(1-2\,x\,\cos{\theta}+x^{2}+\beta^{2})^{1/2}}
\ee
where $\beta$ represents the gravitational softening, as it appeared in the 
WKB approximation. For $\beta>0$, the disk potential and $\Omega$ 
are slightly modified accordingly, but this is a minor effect.}
$\nu_{X}\equiv \partial \ln X/\partial \ln R$, 
and $\Delta \equiv \kappa^{2}-(\Omega-\omega)^{2}$.

\begin{figure}
    \centering
    \scaleup
    \plotonesmall{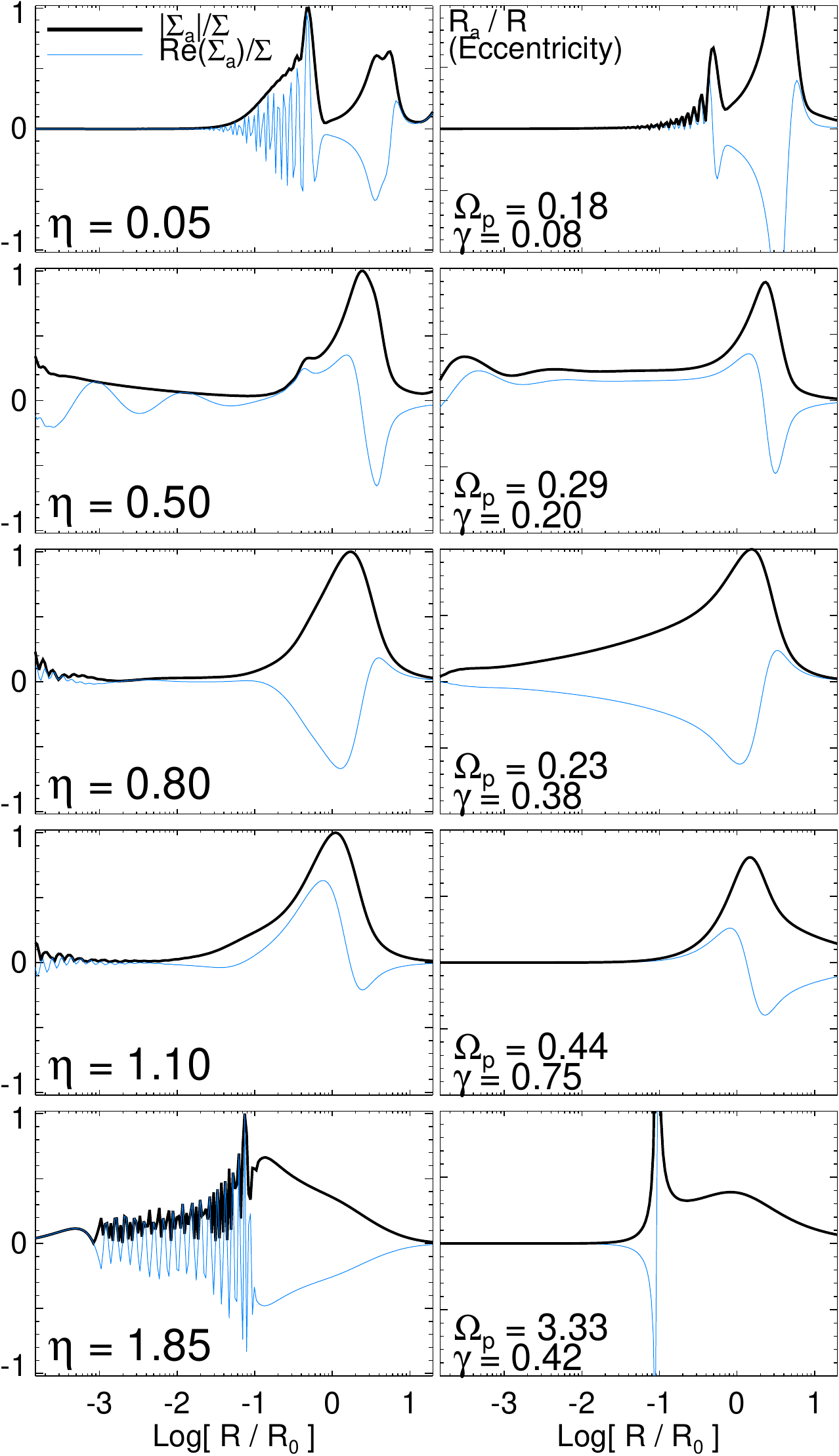}
    \caption{Global linear normal modes for gravitational perturbations to a 
    nuclear stellar disk around a BH; the total disk-to-BH mass ratio $=1$. 
    Each row highlights one mode for a mass profile slope $\Sigma\propto R^{-\eta}$ and
    scale height $\beta\sim h/R=0.1$; the pattern speed $\Omega_{p}$ and 
    growth rate $\gamma$ of the mode are labeled on the right in units of $\Omega(R_0)$.
    Radii are in units of $R_{0}$, the BH radius of influence, $\sim10-100\,$pc. 
    {\em Left:} Mode amplitude:   black is absolute value, blue is ${\rm Re}(\Sigma_{a}/\Sigma)$. 
    {\em Right:} Real and absolute value of the induced eccentricity.
    Shallow surface density profiles ($\eta<1/2$) cannot support modes at small R and so are spatially localized. 
    For larger $\eta$, the modes propagate to $R\rightarrow0$. 
    For $\eta >1$, however, the inwards propagation of eccentricity
    is less efficient (the disk is ``stiffer'' against external perturbations), which would suppress 
    shocks and gas inflow at small $R$. 
    The characteristic profile shapes can therefore self-adjust to have $1/2 \lesssim \eta \lesssim 1$. 
    \label{fig:m1.3}}
\end{figure}


It is straightforward to solve eq. \ref{eqn:mode.eom} for the eigenfunctions (normal modes) of the system. 
For convenience and realism we modify the disk 
mass profiles with a steep outer power-law cutoff 
($\Sigma\propto R^{-\eta}\,(1+[R/a]^{2})^{-(3-\eta)/2}$) 
so that they have finite total mass $M_{d} = M_{\rm BH}$; 
\citet{hopkins:slow.modes} shows that the exact choice of 
$M_{d}$ and/or the cutoff radius has no affect on any of our conclusions. 

Figure~\ref{fig:m1.3} shows some of the resulting normal 
modes for a stellar disk.   The growth rates $\gamma$ and pattern speeds $\Omega_p$ are indicated on each panel, in units of $\Omega(R_0)$.  For any choice of disk parameters, there is a large variety of 
normal modes; here, we focus on the most rapidly growing "global" modes in each model. 
The results are qualitatively similar for all global modes (local modes, 
potentially supported at all radii but localized in radius, are not of interest here). We take $\beta=0.1$ for the softening, but our conclusions are essentially identical for a wide range 
of $\beta$; in \citet{hopkins:slow.modes} we show that this extends to $\beta\gtrsim0.3$, 
i.e.\ nearly-spherical configurations. This is because the manner in which $\Sigma(R)$ 
enters the equations means that the important dimensional parameter is really $M_{\rm enc}(<R)$ 
at a given radius, so puffier systems and even multiple overlapping disks making a quasi-spherical 
configuration give a qualitatively identical result. 
The important parameter we focus on here 
is the power-law index of the disk mass profile $\eta$, for which we 
show various choices in Figure \ref{fig:m1.3}: $\eta=0.05,\,0.50,\,0.80,\,1.10,\,1.85$. 

For each value of $\eta$, Figure~\ref{fig:m1.3} shows the absolute value and real component of the surface density perturbation $a(R)=\Sigma_{a}/\Sigma$ 
and the induced eccentricity $R_{a}/R$, where $R_{a}$ is the magnitude of the radial perturbation 
from the linear equations of motion. The modes are normalized so that ${\rm MAX}(|a(R)|)=1$. 
Where the eccentricities are significant, there can be orbit crossings and  
shocks in the gas.  This dissipation helps drive rapid inflow; see \citet{hopkins:inflow.analytics} for a detailed discussion of this physics.

The key result in Figure~\ref{fig:m1.3}  is how the structure of the global modes changes with 
$\eta$; this confirms our intuition derived from the WKB approximation. 
When the disk surface density profile is shallow ($\eta\lesssim1/2$), the modes 
cannot propagate inwards efficiently -- they are confined to a 
moderate range of radii.   At $\eta=1/2$, the modes are suddenly able propagate to 
arbitrarily small $R$. Going to somewhat larger $\eta=0.7-0.8$, 
the structure of the modes is quite similar. 
For $\eta > 1$, the 
induced eccentricity is strongly suppressed at small radii (see \S~\ref{sec:wkb}) even though the mode 
formally has non-zero amplitude to arbitrarily small $R$.   As a consequence, the induced gas inflow would also be strongly suppressed. 

\vspace{-0.7cm}
\section{Discussion and Conclusions}
\label{sec:discussion}

The general interpretation of the stellar density profiles of $\lesssim L_*$ ellipticals is that 
violent relaxation produces the outer ``wings'' of the mass profile, 
while a nuclear starburst similar to that observed in nearby ULIRGs and many high-redshift galaxies produces a dense stellar relic that dominates the mass profile inside the central $\sim$kpc (e.g., \citealt{faber:ell.centers}).
The mass profile at smaller radii is thus set by the physics of angular momentum redistribution and star formation.   This inner dissipative region is no longer sensitive to large-scale torques such as are produced in a merger.  Instead, the inflow is likely due to secondary gravitational instabilities that develop as the disky component of gas and stars becomes self-gravitating on small scales \citep{shlosman:bars.within.bars}.
  
In previous work \citep{hopkins:zoom.sims} we have demonstrated that inside a radius from one to several times the BH radius of influence, the character of the instabilities that dominate angular momentum redistribution changes: the non-axisymmetry is dominated by an eccentric/lopsided disk or one-armed spiral mode -- a ``slow'' $m=1$ mode, unique to quasi-Keplerian potentials.

If the stellar and gaseous density profiles are relatively shallow (e.g., as might be the case absent an earlier epoch of gas inflow) the $m = 1$ modes cannot be supported down to $R\rightarrow0$, but reflect off of 
a $Q$-boundary at a finite radius.  Between this inner boundary and co-rotation, however, the modes will drive accretion, steepening the gas density profile;  star formation will steepen the corresponding stellar density profile.   As the mass profile steepens, the $m = 1$ modes can propagate deeper in the potential, until a critical slope is reached, at which point the modes can propagate to, and drive inflow to, $R=0$; for a power-law disk with $\Sigma\propto R^{-\eta}$, 
the critical slope is $\eta=1/2$.

If gas is driven to small radii very efficiently, star formation will likely ensure that both the gas and stellar  mass profiles further steepen.   The surface density profile can, however, eventually become sufficiently steep that the inwards propagation of eccentricity is inefficient:  the outer asymmetric perturbation is weak compared to the local disk self-gravity at small radii.  For a power-law disk with $\Sigma\propto R^{-\eta}$, this occurs at $\eta=1$.  The resulting pile-up of mass at larger radii, together with continued star formation, will flatten the gas and stellar mass profiles.

The net result is a plausible equilibrium:   in the presence of significant gas inflow from larger radii, star formation and the propagation of $m = 1$ modes will self-adjust so that the surface density profile 
satisfies $1/2\lesssim\eta\lesssim1$ inside the potential of the central black hole. 
Remarkably, this range of 
slopes is comparable to what is observed in the centers of ``cusp'' ellipticals (Fig. \ref{fig:cusp.examples}).

On the smallest scales near the BH, dynamical relaxation plays an important role in setting the stellar density profile \citep[see][]{bahcallwolf}. Such effects are, however, unlikely to be important 
outside of $\sim1\,$pc because the $N$-body relaxation time becomes 
long compared to the Hubble time.  The presence of massive perturbers can significantly accelerate stellar relaxation, but it is unclear how 
disturbed the nuclei of these (now) gas-poor galaxies are.   Moreover, the observational evidence for disky structures in "cusp" galaxies suggests weak/incomplete 
relaxation, which would have to be very extreme to operate out to $10-100\,$pc in any case.   
Even a merger of binary black holes tends to leave these structures if it is gas-rich \citep{hopkins:zoom.sims}, 
because most of the nuclear gas inflows tend to follow the BH binary coalescence, 
regenerating the disk, and multiple overlapping generations of disks can be formed (which will 
given the same mass profile, since each individually must satisfy the same $\Sigma(R)$ or 
$M_{\rm disk}(R)$ scaling). 
And our conclusions are robust to a number of variations in e.g.\ the disk thickness and detailed structure, 
In the potential of the BH, relaxation can be enhanced by the very same resonance between orbital and epicyclic motion that is so critical for the presence of $m = 1$ modes \citep{kocsis:2010.resonant.relaxation}.   
Scalar resonant relaxation, which modifies the eccentricity axes of stellar orbits, is, however, inefficient at the radii of interest.   Vector resonant relaxation is likely to be important, but this only changes the inclination angles of the disky orbits;  this may wash out some of the observable 
diskyness or introduce warps into the nuclear kinematics, but it will not significantly affect the mass profiles.

The combined effects of star formation and gas inflow driven by $m=1$ modes provide a plausible explanation for the stellar mass profiles of "cusp" ellipticals at $\lesssim 10-100$ pc.  
This explanation links such cuspy profiles to the physics of angular momentum transport and BH growth. It also makes observational predictions: 
(1) the characteristic radii 
of those slopes should be correlated with the radii of influence of the 
BH (or radii enclosing comparable mass); 
(2) non-negligible radial anisotropy should 
be difficult to remove from the stellar orbits; 
(3) similar profile shapes should be observable in late-stage merger remnants, provided sufficient new stars have formed in the central regions.

\acknowledgments   We thank Scott Tremaine for useful conversations.   Support for PFH was provided by the Miller Institute for Basic Research in Science, University of California Berkeley. 

\

\bibliography{/Users/phopkins/Documents/lars_galaxies/papers/ms}

\end{document}